\documentclass[aps,prl,twocolumn,groupedaddress]{revtex4}
\usepackage{epsfig}
\begin{document}

\title{The Anomalous Hall effect in re-entrant ${\bf Au}$Fe  alloys and the real space Berry phase}
\author{ P.Pureur, F. Wolff Fabris, J. Schaf }
\affiliation{Instituto de F\'{i}sica, Universidade Federal do Rio Grande do Sul, \\
Caixa Postal 15051, 91501970 Porto Alegre, RS, Brazil}
\author{ V.N. Vieira}
\affiliation{Instituto de F\'{i}sica e Matem\'{a}tica, Universidade Federal do Pelotas,\\
Caixa Postal 354, 96010900 Pelotas, RS, Brazil}
\author{I.A. Campbell}
\affiliation{Laboratoire des Collo\"{\i}des, Verres, et Nanomateriaux, Universit\'{e} Montpellier II,\\
34095 Montpellier, France}

%\date{\today}

\begin{abstract}

The Hall effect has been studied in a series of ${\bf Au}$Fe samples in the re-entrant concentration range, as well as in the spin glass range. The data demonstrate that the degree of canting of the local spins strongly modifies the anomalous Hall effect, in agreement with theoretical predictions associating canting, chirality and a real space Berry phase. The canonical parametrization of the Hall signal for magnetic conductors becomes inappropriate when local spins are canted.
\end{abstract}

% insert suggested PACS numbers in braces on next line
\pacs{75.10.Nr, 75.50.Lk, 64.60.Fr, 72.10.Fk}

%\maketitle must follow title, authors, abstract, \pacs, 
%and \keywords
\maketitle

The "anomalous" ferromagnetic contribution to the Hall signal was first clearly distinguished by A.W. Smith in 1910 \cite{smith:10}. For many decades, the accepted parametrization of the Hall resistivity in magnetic conductors has been in terms of the canonical expression  
\begin{equation}
\rho_{xy} = R_{h}{\bf B} = R_{0}{\bf B} + R_{s}{\bf M}({\bf B})
\end{equation}
where ${\bf M}({\bf B})$ is the global magnetization, $R_s/\mu_0$ is the anomalous Hall effect (AHE) coefficient and $R_0$ the ordinary (or Lorenz) Hall coefficient. Recently the intrinsic AHE in band ferromagnets has been re-interpreted in terms of the k-space Berry phase \cite{jungwirth:02,onoda:03,fang:03,yao:04,haldane:04}, giving profound new insight into the origin of the Karplus-Luttinger (KL) "anomalous velocity" term \cite{karplus:54} which is usually the major contribution to the AHE. This mechanism leads to a term in $R_{s}$ proportional to the square of the longitudinal resistivity $\rho(T)$, and which otherwise depends only on the band structure and not on the electron scattering. 

But the KL term is not the only contribution to the AHE. In particular for conductors containing spins whose local magnetic axes are tilted away from the global magnetization direction, on theoretical grounds a further AHE term must exist. This is a physically distinct Berry phase contribution occuring in real space when the spin configuration is topologically nontrivial; data on regularly ordered systems such as magnetites and perovskites whose spins are tilted have been interpreted by including this supplementary AHE contribution in the analysis \cite{ye:99,taguchi:01}. The presence of this term is remarkable; it is counter-intuitive because it involves the magnetization components perpendicular to ${\bf M}$. The theoretical principles of this contribution, intrinsically linked to chirality, have now been laid out for the specific case of disordered canted systems such as spin glasses and re-entrant ferromagnets \cite{tatara:02,kawamura:03}. Despite the disorder which at first sight would impose a zero net mean chirality, the coupling between the magnetization and the spin chirality through the spin-orbit interaction leads to a non-zero net chirality when there is a finite magnetization which is either induced by a magnetic field in the spin glass case, or which is spontaneous in the re-entrant case \cite{ye:99,tatara:02,kawamura:03}. However the theory does not for the moment provide a firm order of magnitude or even sign for the effect in specific cases.

We have made systematic measurements of the Hall effect in the $\bf{Au}$Fe alloy series up to $25\%$Fe. The $\bf{Au}$Fe magnetic phase diagram was established by Coles et al \cite{coles:78}, and a wide range of measuring techniques have since been used to study these systems (see \cite{campbell:92} for an overview). Below a critical Fe concentration the alloys are spin glasses, while for higher concentrations the alloys have been dubbed "re-entrant" : on cooling one first encounters a ferromagnetic ordering temperature $T_c$, and then a second canting temperature $T_k$ one of whose signatures is a dramatic drop in the low field ac susceptibility. One now knows that below $T_k$ the system still has an overall ferromagnetic magnetization but the individual Fe spins become canted locally with respect to the global or domain magnetization axis. The drop in susceptibility is due to domain wall pinning through the onset of Dzaloshinski-Moriya interactions when canting sets in \cite{campbell:86,senoussi:88}. For present purposes this alloy series has two main advantages. First, the basic electronic structure of the alloys is that of a noble metal containing  transition metal sites and so can be considered to be relatively simple, in contrast to those of the systems in which chiral AHE effects have been invoked up to now, such as the pyrochlore $Nd_2Mo_20_7$ \cite{taguchi:03} or  magnetites \cite{ye:99}. The resistivity is high in all samples at all temperatures of our study because of impurity scattering \cite{mydosh:74}. With this type of electronic structure, the KL term should be strong and should behave fairly regularly both as a function of temperature and of concentration, and should not be expected to change sign as a function of temperature. Secondly, the degree of canting can be tuned by modifying the Fe concentration $c$. It is well established that as $c$ is increased $T_c$ rises, $T_k$ drops and the degree of canting at low temperatures decreases steadily.  

Hall data measured at lower fields for the region of  $T_g$ in the spin glass alloys ${\bf Au}$Fe and ${\bf Au}$Mn  containing $8\%$ impurity have been analysed to provide evidence for a chiral term \cite{pureur:04,taniguchi:04} . 
The present Hall data demonstrate that there is a strong chiral contribution to the AHE over a wide concentration range covering the spin glass and the re-entrant domains. 

Going beyond the first step which is to verify if the term predicted theoretically is visible experimentally, the ${\bf Au}$Fe alloys will be able to allow very searching tests of  model predictions, because measurements can be made as a function of three control parameters : concentration, temperature, and field. 

We have used standard metallurgical methods to prepare ${\bf Au}$Fe alloys with nominal Fe concentrations of $8, 12, 15, 18, 21$ and $25$ atomic $\%$. Foils were prepared by cold rolling to a thickness of about $20 \mu m$, and were cut into the standard Hall geometry. After cold rolling and cutting the samples were annealed for an hour before quenching. 
Once prepared, the samples were stored in liquid nitrogen to minimize Fe migration effects which can modify the magnetic properties, particularly close to the critical concentration.
For the Hall and resistivity measurements an ac current technique was used having a sensitivity of better than  $10^{-8} V$. 
Fields up to $3T$ could be applied in the Hall geometry at temperatures from $8K$ to room temperature. The magnetization was measured independently at the same fields and temperatures with a commercial Squid magnetometer. The moment values were obtained in low demagnetization factor geometry, and were then corrected appropriately for the Hall geometry demagnetization factor. 

The magnetic properties of the ${\bf Au}$Fe alloys have been studied extensively; see Ref. \cite{coles:78,sarkissian:81,campbell:86,senoussi:88,hennion:86,mirebeau:90,campbell:92, hennion:95} and many other contributions. Up to $13\%$Fe the alloys are spin glasses with the freezing temperature $T_g$ increasing regularly with $c$. From $13\%$Fe to about $30\%$Fe the alloys are re-entrant ferromagnets; the Curie temperature $T_c$ increases strongly with $c$ while the canting temperature $T_k$ decreases steadily towards zero. Neutron diffraction shows that the transverse spin components in the re-entrant phase are not random but that there are transverse ferromagnetic correlations between the spins \cite{hennion:86}; neutron depolarization proves the persistence of ferromagnetic domains down to the lowest temperatures \cite{mirebeau:90}. In this re-entrant region the usual canting temperature $T_k$ estimates which we quote correspond to static or low frequency measurements, but for temperatures between $T_c$ and $T_k$ inelastic neutron diffraction (which is a high frequency measurement) shows magnon softening indicating a slowing down of canting dynamics \cite{hennion:95}. 

The Hall coefficent $R_{h}(T)$ is shown in figure 1 for four alloys as a function of temperature at a single fixed magnetic field. For this figure the fields have been chosen such that the re-entrant samples are close to  technical saturation at low $T$. On the same plots we show a calculated $R_{h}^{*}(T)$
estimated assuming only the canonical contributions :
\begin{equation}
R_{h}^{*}(T) = R_0(c) + A(c)M_{h}(T)[\rho(T)]^2
\end{equation}
where the first term is the ordinary Hall coefficient and the second term is an estimate of the KL AHE contribution calculated using $M_{h}(T)$ (the measured ratio of the magnetization in Hall geometry to the applied field) times the square of the resistivity $\rho(T)$ multiplied by a concentration dependent constant $A(c)$. $R_{0}(c)$ and $A(c)$ were estimated by first plotting $R_{h}(T)$ against $M_{h}(T)[\rho(T)]^2$. For each sample the data in the high temperature range fall on a straight line, which is consistent with the assumption that the these conventional terms dominate at high  $T$. The intercept and the slope of each  line provide us with values of $R_0(c)$ and $A(c)$ respectively in equation 2.  In these concentrated alloys and at the fields indicated $R_0$ makes only  a small relative contribution to the total $R_{h}$ except towards the very high temperature limit. For the lower concentrations its value is close to that of Au metal, $-7.10^{-11} m^3/C$ \cite{hurd:72} but $R_0(c)$ then evolves towards positive values, changing sign near 13$\%$ Fe (c.f. \cite{mcalister:76,barnard:88}). For the concentrations for which we show data,  $A(c)$ can be estimated accurately from the $R_{h}(T)$ against $M_{h}(T)[\rho(T)]^2$ plots. $A(c)$ evolves from negative at low Fe concentrations to positive at high concentrations with a change of sign near $16\%$Fe. (This behaviour is very similar to that of the ferromagnetic  ${\bf Ni}$Fe and ${\bf Pd}$Fe alloy series which can be expected {\it a priori} to have a broadly similar electronic structures to the ${\bf Au}$Fe series;  the AHE exponent $R_s(c)$ passes from negative to positive near 13$\%$Fe in ${\bf Ni}$Fe \cite{jellinghaus:60} and near 18$\%$Fe in ${\bf Pd}$Fe \cite{abramova:74}).  

A "conventional" AHE $R_h ^{*}(T)$ was calculated over the entire temperature range assuming  $A(c)$ and $R_0$ to be temperature independent. In fact this is an approximation; in particular once in the ferromagnetic regime $R_0$ depends on the effective spin-up to spin-down resistivity ratio \cite{campbell:70} and so will be temperature dependent.  For the data reported here this should represent a minor correction. On the other hand as the basic electronic structure of these alloys is simple (in contrast to that of the ferromagnetic perovskite $SrRuO_3$ for instance \cite{fang:03,kats:04}) one should expect that a temperature independent $A(c)$ in the KL term for each alloy should be a reasonable approximation. It can be noted that below $T_c$ the absolute value of the calculated $R_{h}^{*}(T)$ tends to drop in the re-entrant alloys because as $T$  decreases the drop in resistivity more than compensates the increase of magnetization. 

Except for the  $25\%$Fe sample there is a striking difference between the measured $R_{h}(T)$ and the $R_{h}^{*}(T)$ curve calculated with the conventional contributions only, leading to a total  $R_{h}(T)$ which changes sign with temperature for the intermediate concentrations. 

We ascribe the difference $[R_h(T)-R_h^*(T)]$ to the canting term. Consider first the high concentration end at $25\%$Fe, where we know that low temperature canting is weak (but not zero \cite{rakoto:84}). $R_h$ is dominated over the whole temperature range by the  KL term, positive at this concentration; the measured $R_h(T)$ is only slightly less positive than the calculated $R_h^*(T)$ for temperatures below $T_c$. As the concentration is then reduced step by step towards and beyond the critical concentration of $13\%$Fe, we know that the low temperature canting of the spins progressively increases. In the limit $T$ tending to zero, a negative $[R_h(0)-R_{h}^{*}(0)]$ term steadily develops for the sequence of alloys $21\%$ Fe, $18\%$ Fe, $15\%$ Fe, $12\%$Fe. For the first two alloys there is a change of sign with temperature (c.f. \cite{hurd:79}), and for the $15\%$ alloy where the KL term is weak ($15\%$Fe is close to the concentration where this term changes sign) the negative term dominates over almost the entire temperature range. Finally when we pass the transition into the spin glass alloy region, for the $12\%$Fe there remains a negative contribution with respect to the calculated $R_{h}^{*}(T)$  which peaks in the neighbourhood of $T_g$. The difference term then becomes positive by $8\%$Fe as was observed at lower applied fields \cite{pureur:04}. The change in sign in the canting term may be associated with the difference between ferromagnetic correlations among the canted spin components for the more concentrated alloys and quasi-random correlations well in the spin glass region. 

\begin{figure}
\includegraphics[width=9cm, height=13cm,angle=0]{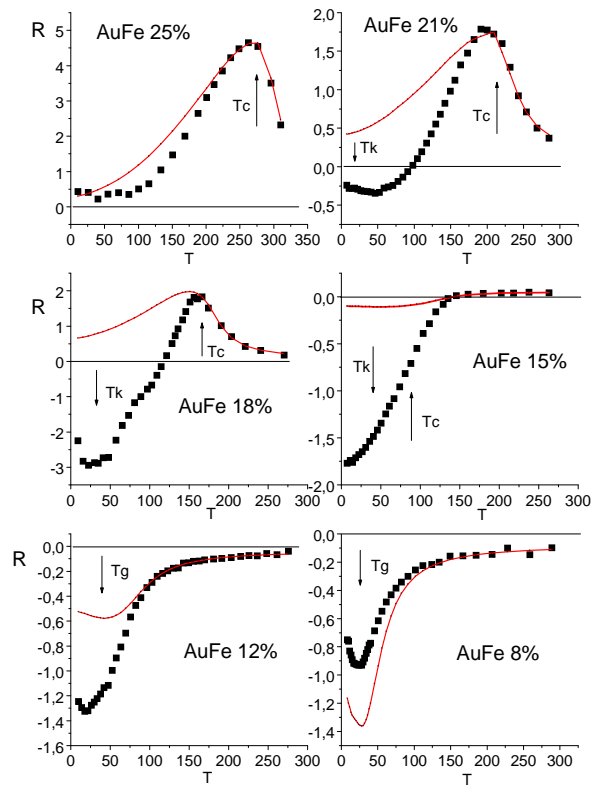}
\caption{The total Hall effect $R_{h}(T)$ measured in the six samples of the ${\bf Au}$Fe alloy series as functions of temperature, at applied fields of  $0.5 T$ for the two highest concentrations, and of  $0.25T$ for the others.  $R_h$ is measured in units of $10^{-9}m^3/C$. For each panel $R_{h}^{*}(T)$ is calculated for the same field assuming that only the standard ordinary Hall and KL terms contribute  (see equation (2)). $R_{h}^{*}(T)$  is shown as a continuous line. Concentrations are listed going downwards from $25\% Fe$ to $8\% Fe$.  Ordering temperatures $T_c$ and $T_g$ were measured on the present samples using low field magnetization data; canting temperatures $T_k$ are quoted from Coles et al \cite{coles:78}. } 
\protect\label{fig:DS}
\end{figure}

The chiral or real space Berry phase theory as applied to the re-entrant systems \cite{tatara:02,kawamura:03} states that local spin canting should lead to the novel chiral AHE term, but makes no firm prediction concerning either the sign or the strength of the effect. The present experimental data in the low temperature limit demonstrate that in the re-entrant alloys there is indeed a large positive contribution over and above the canonical KL term, and that the strength of this contribution is closely correlated with the degree of canting. At low temperatures, this term is large enough to dominate the KL term over almost all the re-entrant region . Because of the clear correlation with the presence of canting the difference term can be confidently identified with the theoretically predicted chiral or real space Berry phase term. 

Once this point established, we can discuss the temperature dependence of the effect. At face value, in the re-entrant concentration range the theory \cite{kawamura:03} would predict an onset of the chiral term $R_h(T)-R_{h}^{*}(T)$ only below the static canting temperature $T_k(c)$. The data in Figure 1 indicate inequivocally that the extra term appears already at temperatures well above $T_k(c)$ for each concentration. This can be understood at least at the phenomenological level by taking into account the relatively slow relaxation of the transverse $[x,y]$ components of the spins even above $T_k(c)$. One can conjecture that the Berry phase Hall effect like a normal transport property is sensitive to magnetic configurations on a time scale of the electron scattering time, i.e. typically $10^{-12}$ seconds. When the local $[x,y]$ component relaxation is slower than this, the conduction electron will sense the canting as frozen. Hence the canting AHE term can be expected to appear at some temperature, higher than $T_k$, whose value will be controled by the local spin relaxation rate. Neutron scattering and $ \mu SR$ data give indications of this relaxation rate \cite{hennion:95,pappas}. A quantitative discussion must await a detailed analysis of the relaxation data, and a more complete theoretical understanding of the real space Berry phase mechanism in the presence of relaxing spins would be very welcome. To give a quantitative analysis of the data it would also be necessary to have in hand model predictions for the effect of spatial correlations between the $[x,y]$ components on the strength of the Berry phase term. 
(Though Taniguchi et al \cite{taniguchi:04} also appeal to the same theoretical models, their interpretation of Hall data an alloy similar to ours is entirely different as they consider that only the differences between  $R_h$  values after Zero Field Cooled and Field Cooled protocols can be ascribed to the canting term).

We conclude that the chiral or Berry phase term should thus be present in any conductor containing statically canted local spins, and also  in conductors with spins which are effectively canted when their relaxation rate is smaller than the conduction electron scattering rate even if they are aligned ferromagnetically on average over long time scales. In these circumstances equation (1) can still be written down formally, but it loses all transparency because physical phenomena depending not only on the bulk magnetization but on the details of the transverse local spin structure and its dynamics will be hidden within the AHE parameter $R_s$. There should at least be a further term depending on $[<\mid M_{\perp}\mid>] $, the average local magnetization component perpendicular to the global magnetization axis, and even this modification would be insufficiant as a full interpretation of the Hall signal would require detailed knowledge of the microscopic magnetic configuration of the local spins, and of their dynamics. 

In summary, we have presented experimental AHE data for the re-entrant ${\bf Au}$Fe alloy series which demonstrate conclusively the presence of a strong AHE term linked to local spin canting, providing clear experimental evidence which supports rigorous but qualitative theoretical predictions based on chirality or real space Berry phase considerations \cite{ye:99,tatara:02,kawamura:03}. This mechanism has an entirely different physical origin from that of other contributions involved in the interpretation of the AHE, and the present results show that it can be important even in metals with relatively simple band structures. The influence of spin dynamics on this AHE term does not seem to have been considered up to now.  

We would like to thank  Dr. G. Tatara and Professor H. Kawamura for helpful discussions.

\end{document}